\documentclass[article]{./macros/elsarticle_qh}

\usepackage[hmarginratio=1:1,top=32mm,left=20mm,columnsep=1cm]{geometry} %
\usepackage{bm}
\usepackage{amsmath}
\usepackage{relsize} 
\usepackage[svgnames]{xcolor} 
\usepackage{grffile} 
\usepackage{graphicx}

\usepackage{hyperref}
\hypersetup{
	colorlinks=true,                          
	linkcolor=DarkRed,
	citecolor=DarkRed,
	urlcolor=DarkRed  }

\usepackage{bm}
\usepackage{amsmath}
\usepackage{booktabs}
\usepackage{amssymb}
\usepackage{etoolbox}
\usepackage{mathtools}
\usepackage{gensymb}
\usepackage[section]{placeins} 
\usepackage{float}
\usepackage{subfig}

\usepackage[thinc]{esdiff}

\DeclareMathAlphabet{\mathsfbi}{OT1}{\sfdefault}{bx}{sl}
\DeclareMathVersion{sfletters}
\SetSymbolFont{letters}{sfletters}{OML}{ntxsfmi}{b}{it}

\makeatletter
\newcommand{\mathbfsbilow}[1]{%
	\text{\mathversion{sfletters}$\m@th#1$}%
}
\DeclareRobustCommand{\tensor}[1]{%
	\begingroup
	\ifcat\noexpand #1\relax
	\edef\greek@test{\detokenize{#1}}%
	\edef\greek@test{\expandafter\@cdr\greek@test\@nil}%
	\edef\greek@test{\expandafter\@car\greek@test\@nil}%
	\edef\x{\the\lccode\expandafter`\greek@test}%
	\edef\y{\number\expandafter`\greek@test}%
	\ifnum\x=\y\relax
	\mathbfsbilow{#1}%
	\else
	\mathsfbi{#1}%
	\fi
	\else
	\mathsfbi{#1}%
	\fi
	\endgroup
}
\makeatother
\makeatletter
\newcommand{\sbullet}{%
	\hbox{\fontfamily{lmr}\fontsize{.4\dimexpr(\f@size pt)}{0}\selectfont\textbullet}}

\makeatother

\usepackage[numbers]{natbib}

\begin{document}
	
\begin{frontmatter}
		
\title{{\Large \textbf{Thomsen-type parameters and attenuation coefficients for constant-\textit{Q} transverse isotropy}}
}

\cortext[mycorrespondingauthor]{Corresponding author}

\address[JLU]{College of Geoexploration Science and Technology, Jilin University, Changchun, 130026, P. R. China}
\address[CSM]{Department of Geophysics, Colorado School of Mines, Golden, 80401, USA}

\author[JLU]{Qi Hao\corref{mycorrespondingauthor}} 
\ead{{xqi.hao@gmail.com}}

\author[CSM]{Ilya Tsvankin}
\ead{ilya@mines.edu}

\begin{abstract}
Transversely isotropic (TI) media with the frequency-independent quality-factor elements (also called ``constant-$Q$'' transverse isotropy) are often used to describe attenuation anisotropy in sedimentary rocks. 
The attenuation coefficients in constant-$Q$ TI models can be conveniently defined in terms of the Thomsen-type attenuation-anisotropy parameters. Recent research indicates that not all those parameters for such constant-$Q$ media are frequency-independent. Here, we present concise analytic formulae for the Thomsen-type attenuation parameters for Kjartansson's constant-$Q$ TI model and show that one of them ($\delta_{Q}$) varies with frequency. The analytic expression for $\delta_{Q}$ helps evaluate the frequency dependence of the normalized attenuation coefficients of P- and SV-waves by introducing the newly defined ``dispersion factors''. Viscoacoustic constant-$Q$ transverse isotropy is also discussed as a special case, for which the elliptical condition and simplified expressions for the parameters $\delta$ and $\delta_{Q}$ are derived. Our results show that in the presence of significant absorption the attenuation coefficients of the ``constant-$Q$'' model vary with frequency for oblique propagation with respect to the symmetry axis. This variation needs to be taken into account when applying the spectral-ratio method and other attenuation-analysis techniques. 
\end{abstract}


\end{frontmatter}

\section{Introduction}
The frequency-independent quality factor (called ``constant-$Q$'' for brevity)  provides a useful phenomenological description of seismic attenuation in rocks and is widely used in seismic attenuation analysis. Among such constant-$Q$ dissipative models are those proposed by \cite{kolsky:1956} and \cite{kjartansson:1979}. For isotropic media, the Kjartansson model produces the constant-$Q$ factor for all frequencies, whereas the Kolsky model leads to nearly constant $Q$-values. The complex moduli for the Kolsky model represent the first-order Maclaurin series expansion with respect to $1/Q$ of the corresponding moduli for the Kjartansson model \cite[]{hao.greenhalgh:2021a,hao.greenhalgh:2021b}. 

As an extension of non-dissipative transverse isotropy, the constant-$Q$ TI model can be used to process seismic attenuation data for most sedimentary rocks, such as shale formations.  The constant-$Q$ assumption facilitates estimation of the quality factor and  attenuation anisotropy \citep[e.g.,][]{behura:2009,shekar:2011,shekar:2012,behura:2012}. Ultrasonic measurements demonstrate that attenuation anisotropy generally is stronger than velocity anisotropy for rock samples \cite[]{best:2007,zhu.tsvankin:2007b,zhubayev:2016}. 

Velocity anisotropy for TI media can be efficiently described by the Thomsen anisotropy parameters \cite[]{thomsen:1986,tsvankin:2001}. Likewise, attenuation anisotropy for dissipative TI media is convenient to study using the Thomsen-type notation introduced by \cite{zhu.tsvankin:2006}.  The combination of the velocity- and attenuation-related Thomsen-type parameters \cite[]{zhu.tsvankin:2006} completely defines the complex stiffness matrix at a specified frequency for a general dissipative TI model with a vertical symmetry axis (VTI medium). The generic Thomsen velocity parameters depend on the real parts of the stiffness coefficients ($c_{ij}$), whereas the Thomsen-type attenuation parameters are defined by both the real and imaginary parts of $c_{ij}$.  \cite{hao.greenhalgh:2022b} found that some Thomsen-type parameters in constant-$Q$ VTI media are frequency dependent. This phenomenon is not entirely surprising, because all stiffness coefficients of constant-$Q$ TI media, which are involved in the definition of the Thomsen-type parameters, are functions of frequency. Investigating the frequency variations of these parameters can facilitate understanding of such key signatures in TI media as velocities, traveltimes, attenuation coefficients, and polarization vectors. However, to our knowledge, there are no  analytic expressions for the frequency-dependent Thomsen-type attenuation parameters in constant-$Q$ dissipative VTI media. 

Here, we derive analytic formulae for the Thomsen-type parameters using Kjartansson's constant-$Q$ VTI model. A set of the corresponding reference parameters is defined at a specified frequency and used to obtain those parameters for the entire frequency range. We also present a formula for the frequency-dependent anellipticity and define the condition for elliptical anisotropy in constant-$Q$ TI media. The newly proposed formulae for the Thomsen-type parameters allow us to study the normalized plane-wave attenuation coefficients in constant-$Q$ media with weak attenuation anisotropy and define the ``dispersion factors'' for P- and SV-waves. Numerical examples are used to analyze the accuracy of the obtained expressions for the Thomsen-type parameters, the validity of the elliptical condition, and the frequency dependence of the attenuation coefficients. 

\section{Thomsen-type parameters of constant-Q VTI media}
\subsection{Constant-Q dissipative VTI model}
Referring to \cite{zhu.tsvankin:2006} and \cite{carcione:2014}, the complex stiffness (or modulus) matrix $ \mathbf{M}$ for viscoelastic VTI media is given by:
\begin{equation} \label{eq:Mgen}
\small{
\mathbf{M} =
\left(
\begin{matrix}
M_{11} & M_{11}-2M_{66} & M_{13} & 0 & 0 & 0 \\
M_{11}-2M_{66} & M_{11} & M_{13} & 0 & 0 & 0 \\
M_{13} & M_{13} & M_{33} & 0 & 0 & 0 \\
0              & 0              & 0              & M_{55} & 0 & 0 \\
0              & 0              & 0              & 0 & M_{55} & 0 \\
0              & 0              & 0              & 0 & 0 & M_{66}
\end{matrix}
\right),
}
\end{equation}
where $M_{ij}=M_{ij}^R - i \,\text{sgn}(f) M_{ij}^{I}$ denote the complex stiffness coefficients for the frequency $f$, and the minus sign in front of $i \, \text{sgn}(f) M_{ij}^{I}$ follows from the definition of the Fourier transform in \cite{cerveny.psencik:2009} and \cite{hao.greenhalgh:2021a}. Both the real and imaginary parts of $M_{ij}$ generally are frequency dependent. 

For the \cite{kjartansson:1979} model (also called the constant-$Q$ model), the nonzero independent elements in equation \ref{eq:Mgen} are expressed as:
\begin{equation}  \label{eq:Mkjar}
	M_{ij} = \frac{\tilde{M}^{R}_{ij}}{\text{cos}(\pi \gamma_{ij}) }
 \left(-i\frac{f}{f_{0}} \right)^{2\gamma_{ij}} ,
\end{equation}
with
\begin{equation} \label{eq:gamij}
	\gamma_{ij} = \frac{1}{\pi} \, \text{tan}^{-1} \left( \frac{1}{Q_{ij}} \right) ,
\end{equation}
where $Q_{ij} \equiv M_{ij}^R/M_{ij}^{I}$, $f_{0}$ is the reference frequency, and $\tilde{M}^{R}_{ij}$ denote the real parts of $M_{ij}$ at $f_{0}$:  $\tilde{M}^{R}_{ij}=\text{Re}(M_{ij})|_{f=f_0}$.  By design, the quality-factor elements $Q_{ij}$ for the Kjartansson model are independent of frequency. As follows from equation \ref{eq:Mkjar}, the complex stiffness coefficients $M_{ij}$ for a given frequency can be expressed in terms of $\tilde{M}_{ij}^R$ and $Q_{ij}$. 

\subsection{Thomsen-type parameterization}
\cite{zhu.tsvankin:2006} and \cite{tsvankin.grechka:2011} show that dissipative VTI media can be conveniently parameterized by the Thomsen-type attenuation parameters. The \cite{thomsen:1986} velocity parameters  \citep[see][]{tsvankin:2001} are defined in the nonattenuative reference VTI medium. 

The parameter $V_{P0}$ is the vertical velocity of P-waves: 
\begin{equation} \label{eq:vp0def}
V_{P0}\equiv \sqrt{\frac{M_{33}^{R}}{\rho}} ,
\end{equation} 
where $\rho$ denotes density. 

The parameter $V_{S0}$ is the vertical velocity of S-waves:
\begin{equation} \label{eq:vs0def}
V_{S0} \equiv \sqrt{\frac{M_{55}^{R}}{\rho}} .
\end{equation} 

The parameter $\epsilon$ is approximately equal to the fractional difference between the horizontal and vertical velocities of P-waves:
\begin{equation} \label{eq:epsdef}
\epsilon \equiv \frac{M_{33}^{R} - M_{11}^{R}}{2M_{33}^{R}} .
\end{equation}

The parameter $\delta$ determines the second derivative of the P-wave phase velocity at vertical incidence and is given by: 
\begin{equation} \label{eq:deltadef} 
\def \tmpa { \left( M_{13}^{R} +M_{55}^{R}  \right)^2 }
\def \tmpb { \left( M_{33}^{R} -M_{55}^{R}  \right)^2  }
\def \tmpc { 2M_{33}^{R} (M_{33}^{R} - M_{55}^{R}) }
\delta  \equiv \frac{\tmpa - \tmpb}{\tmpc}  .
\let \tmpa \undefined
\let \tmpb \undefined
\let \tmpc \undefined
\end{equation}

The parameter $\gamma$ is approximately equal to the fractional difference between the horizontal and vertical velocities of SH-waves: 
\begin{equation} \label{eq:gamdef}
\gamma \equiv \frac{M_{66}^{R} - M_{55}^{R}}{2M_{55}^{R}} .
\end{equation}

The Thomsen-type attenuation parameters \cite[]{zhu.tsvankin:2006} can be used to define the normalized phase attenuation coefficient $\mathcal{A} \equiv |\mathbf{k}_{I}|/|\mathbf{k}_{R}|$ for P-, SV-, and SH-waves, which is generally supposed to be frequency-independent in constant-$Q$ models. For more details about $\mathcal{A}$, see the section ``Plane-wave attenuation in constant-$Q$  VTI media'' below.

The parameter $\mathcal{A}_{P0}$ is the vertical attenuation coefficient of P-waves:
\begin{equation} \label{eq:Ap0def}
\mathcal{A}_{P0} \equiv Q_{33} \left( \sqrt{1 + \frac{ 1 }{Q_{33}^2}} - 1 \right) \approx 
\frac{ 1 }{2Q_{33}} .
\end{equation} 

The parameter $\mathcal{A}_{S0}$ is the vertical attenuation coefficient of S-waves: 
\begin{equation} \label{eq:As0def}
\mathcal{A}_{S0} \equiv Q_{55} \left( \sqrt{1 + \frac{ 1 }{Q_{55}^2}} - 1 \right) \approx
\frac{ 1 }{2Q_{55}} .
\end{equation} 

The parameter  $\epsilon_{Q}$ is close to the fractional difference between the horizontal and vertical attenuation coefficients of P-waves: 
\begin{equation} \label{eq:epsQdef}
\epsilon_{Q} \equiv \frac{Q_{33} - Q_{11}}{Q_{11}} .
\end{equation}

The parameter $\delta_{Q}$ controls the second derivative of the P-wave  attenuation coefficient at vertical incidence and is expressed as \cite[]{zhu.tsvankin:2006}: 
\begin{equation}   \label{eq:deltaQdef}
\def \tmpa { \frac{Q_{33}-Q_{55}}{Q_{55}} }
\def \tmpb { M_{55}^{R} \frac{\left( M_{13}^{R}
+ M_{33}^{R} \right)^{2}}{M_{33}^{R}-M_{55}^{R}} }
\def \tmpc { 2\frac{Q_{33}-Q_{13}}{Q_{13}} }
\def \tmpd { M_{13}^{R} (M_{13}^{R} + M_{55}^{R}) }
\def \tmpe { M_{33}^{R} (M_{33}^{R} - M_{55}^{R}) }
\delta_{Q}  \equiv \frac{\tmpa \tmpb + \tmpc \tmpd}{\tmpe}  .
\let \tmpa \undefined
\let \tmpb \undefined
\let \tmpc \undefined
\let \tmpd \undefined
\let \tmpe \undefined
\end{equation}

The parameter  $\gamma_{Q}$ is close to the fractional difference between the horizontal and vertical  attenuation coefficients of SH-waves: 
\begin{equation} \label{eq:gamQdef}
\gamma_{Q} \equiv \frac{Q_{55} - Q_{66}}{Q_{66}} .
\end{equation}

\section{Analytic description of Thomsen-type parameters}
In this section, we represent the Thomsen velocity parameters and Thomsen-type attenuation parameters in terms of their reference values defined at $f=f_{0}$:   $\tilde{V}_{P0}=V_{P0}|_{f=f_0}$, $\tilde{V}_{S0}=V_{S0}|_{f=f_0}$, $\tilde{\epsilon}=\epsilon|_{f=f_0}$, $\tilde{\delta}=\delta|_{f=f_0}$, $\tilde{\mathcal{A}}_{P0}=\mathcal{A}_{P0}|_{f=f_0}$, $\tilde{\mathcal{A}}_{S0}=\mathcal{A}_{S0}|_{f=f_0}$, $\tilde{\epsilon}_{Q}=\epsilon_{Q}|_{f=f_0}$, and $\tilde{\delta}_{Q}=\delta_{Q}|_{f=f_0}$. These parameters are used to find the real parts of the reference stiffness coefficients ($\tilde{M}_{ij}^R$), the quality-factor elements $Q_{ij}$ (see Appendix A), and the frequency-dependent stiffness matrix $\mathbf{M}$. 

According to equations \ref{eq:vp0def}--\ref{eq:deltadef} and \ref{eq:deltaQdef}, the Thomsen-type parameters involve the coefficients $M_{ij}^R$, where $ij$=11, 13, 33, 55 and 66. Using equations \ref{eq:Mkjar} and \ref{eq:gamij},   $M_{ij}^R$ are approximately expressed as:
\begin{equation} \label{eq:Mijappr}
M_{ij}^R \approx \tilde{M}_{ij}^R \left(
1 + \frac{2}{\pi} \, Q_{ij}^{-1} \ln{\bigg| \frac{f}{f_{0}} \bigg| }
+ \frac{2}{\pi^2} \, Q_{ij}^{-2} \ln^2{ \bigg| \frac{f}{f_{0}} \bigg| }
\right) ,
\end{equation}
where we use the approximation $\text{tan}^{-1}(Q_{ij}^{-1}) \approx Q_{ij}^{-1}$ because typically $Q_{ij} \gg 1$. 

Substitution of equation \ref{eq:Mijappr} into equations \ref{eq:vp0def}--\ref{eq:deltadef} and  \ref{eq:deltaQdef} allows us to derive approximate expressions for the frequency-dependent Thomsen-type parameters, which are discussed in the following two subsections. 

\subsection{Velocity parameters}
The second-order approximations for the Thomsen velocity parameters with respect to $\text{ln} |f/f_{0}|$ are given by:
\begin{equation} \label{eq:vp0f}
V_{P0} = \tilde{V}_{P0}   \left( 1 + 
\frac{1}{\pi} \, Q_{33}^{-1} \ln{\bigg| \frac{f}{f_{0}} \bigg|}
+ \frac{1}{2\pi^2} \, Q_{33}^{-2} \ln^2{\bigg| \frac{f}{f_{0}} \bigg| } 
\right) ,
\end{equation}
\begin{equation} \label{eq:vs0f}
V_{S0} = \tilde{V}_{S0}   \left( 1 + 
\frac{1}{\pi} \, Q_{55}^{-1} \ln{\bigg| \frac{f}{f_{0}} \bigg|}
+ \frac{1}{2\pi^2} \, Q_{55}^{-2} \ln^2{\bigg| \frac{f}{f_{0}} \bigg| } 
\right) ,
\end{equation}
\begin{align}
\label{eq:epsf}
&\epsilon =  \tilde{\epsilon}  
+ \frac{1}{\pi}  (1+2\tilde{\epsilon}) \, Q_{33}^{-1} \, \tilde{\epsilon}_{Q} \ln{\bigg| \frac{f}{f_{0}} \bigg|}
+ \frac{1}{\pi^2} (1+2\tilde{\epsilon}) \, Q_{33}^{-2} \, \tilde{\epsilon}_{Q}^2 
 \ln^2{\bigg| \frac{f}{f_{0}} \bigg| } \, , \\
\label{eq:deltaf}
&\delta = \tilde{\delta} 
+ \frac{1}{\pi} \, Q_{33}^{-1} \, \tilde{\delta}_{Q} \ln{\bigg| \frac{f}{f_{0}} \bigg|}
+  \frac{1}{\pi^2} \, Q_{33}^{-2} \, \zeta_{Q} 
 \ln^2{\bigg| \frac{f}{f_{0}} \bigg| } \, ,  \\
\label{eq:gamf}
&\gamma = \tilde{\gamma}  
+ \frac{1}{\pi} (1+2\tilde{\gamma}) \, Q_{55}^{-1} \, \tilde{\gamma}_{Q} \ln{\bigg| \frac{f}{f_{0}} \bigg|}
+ \frac{1}{\pi^2} (1+2\tilde{\gamma}) \, Q_{55}^{-2} \, \tilde{\gamma}_{Q}^2 
 \ln^2{\bigg| \frac{f}{f_{0}} \bigg| } \, , 
\end{align}
where the P- and S-wave inverse vertical quality factors $Q_{33}$ and $Q_{55}$ (respectively) are:
\begin{align}
\label{eq:Qp0}
&Q_{33} ^{-1}= \frac{2 \tilde{\mathcal{A}}_{P0}}{1 - \tilde{\mathcal{A}}_{P0}^2} , \\
\label{eq:Qs0}
&Q_{55}^{-1} = \frac{2 \tilde{\mathcal{A}}_{S0}}{1 - \tilde{\mathcal{A}}_{S0}^2} . 
\end{align}
The coefficient $\zeta_{Q}$ in equation \ref{eq:deltaf} is defined as:
\begin{equation} \label{eq:zetaQ}
\zeta_{Q} = 
d_{0} \left(1 - g_{Q} \right)^2 
+ d_{1} \left( 1- g_{Q}  \right) \tilde{\delta}_{Q}   
+ d_{2} \, \tilde{\delta}_{Q}^2 \, ,
\end{equation}
with
\begin{equation} \label{eq:gQ}
g_{Q} \equiv \frac{Q_{33}}{Q_{55}} , 
\end{equation}
\begin{align}
\label{eq:d0}
&d_{0} =\frac{
g(1-g+\chi)^2 \left[ (1+2\tilde{\delta})\, \chi - (1+2\tilde{\delta})\, g + (1+\tilde{\delta})\, g^2 \right] 
}{ (1-g)^2 (\chi-g) \chi^2 } , \\
\label{eq:d1}
&d_{1} =\frac{ 2g\left[ 1+2\tilde{\delta}+\chi - (2+\tilde{\delta}+\chi) \, g + g^2 \right] }{ (\chi - g)\chi^2 } , \\
\label{eq:d2}
&d_{2}=\frac{ 2\chi - g }{2(1+2\tilde{\delta} - g)(\chi - g) } \, ;
\end{align}
\begin{align}
\label{eq:g}
&g \equiv \frac{\tilde{V}_{S0}^2}{\tilde{V}_{P0}^2} , \\
\label{eq:chi}
&\chi=\sqrt{(1-g)(1+2\tilde{\delta} - g)} \, .
\end{align}

In equations \ref{eq:vp0f}--\ref{eq:gamf}, the first-order terms with respect to $\text{ln} |f/f_{0}|$ are scaled by $Q_{33}^{-1}$ or $Q_{55}^{-1}$, whereas the second-order terms by $Q_{33}^{-2}$ or $Q_{55}^{-2}$. Because $Q_{33}$ and $Q_{55}$ typically are much greater than unity, the frequency dependence of the velocity parameters is mostly determined by the first-order terms.  Equations \ref{eq:vp0f}--\ref{eq:gamf} indicate that: (1) $V_{P0}$ and $V_{S0}$ always monotonically increase with frequency; (2) $\epsilon$, $\delta$ and $\gamma$ also monotonically increase with $f$, if $\tilde{\epsilon}_{Q}>0$, $\tilde{\delta}_{Q}>0$, and $\tilde{\gamma}_{Q}>0$, respectively. Overall, the frequency dependence of $V_{P0}$, $V_{S0}$, $\epsilon$, $\delta$, and $\gamma$ for realistic values of $Q_{33}$ and $Q_{55}$ ($Q_{33} \gg 1$ and $Q_{55} \gg 1$) remains weak, as illustrated by the numerical examples below.  

\begin{figure}[H]
	\centering
	\includegraphics[width=2.8in]{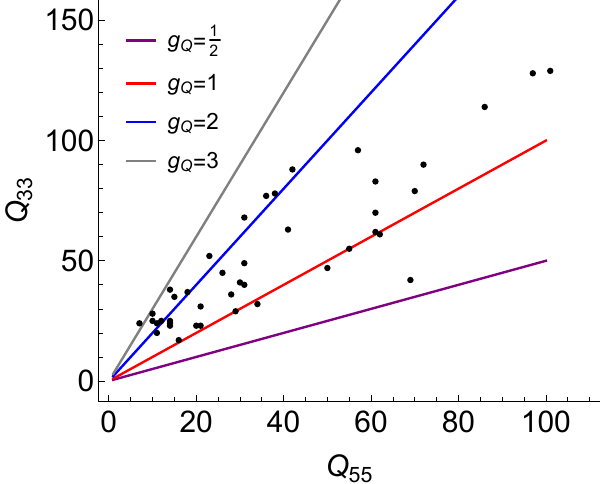}
	\caption{
     Vertical quality factors $Q_{33}$ and $Q_{55}$ in dissipative VTI rocks. The black dots are the data from Table 3 of \cite{best:2007}; $g_{Q} \equiv Q_{33}/Q_{55}$. 
	}
	\label{fig:gQ_rocks}
\end{figure}

Note that phase and group velocities in strongly dissipative TI media are influenced by attenuation and do not represent the same functions of the Thomsen parameters as in purely elastic models \cite[]{zhu.tsvankin:2006,tsvankin:2001}. For sedimentary formations, both $g_{Q}$ and $g$ vary within a limited range. In particular, according to \cite{best:2007}, for relatively shallow sedimentary rocks $ 0.5 < g_{Q} \le 3$ (Figure \ref{fig:gQ_rocks}). 

Using equations \ref{eq:epsf} and \ref{eq:deltaf} for $\epsilon$ and $\delta$, the anellipticity parameter $\eta$ \cite[]{alkhalifah:1995} can be approximately obtained as:
\begin{equation} \label{eq:etaf}
\eta \equiv \frac{ \epsilon-\delta }{ 1 + 2\delta } = \eta_{0} +  \eta_{1} \, Q_{33}^{-1} \ln{\bigg| \frac{f}{f_{0}} \bigg|} + \eta_{2} \, Q_{33}^{-2} \ln^2{\bigg| \frac{f}{f_{0}} \bigg|} ,
\end{equation}
where $ Q_{33}$ is given by equation \ref{eq:Qp0}, and
\begin{align}
\label{eq:eta0}
&\eta_{0} = \tilde{\eta} = \frac{\tilde{\epsilon}-\tilde{\delta}}{1 + 2\tilde{\delta}} ,  \\
\label{eq:eta1}
&\eta_{1}=\frac{1 + 2\tilde{\epsilon}}{(1 + 2\tilde{\delta})^2}
\left[ (1+2\tilde{\delta})\tilde{\epsilon}_{Q}-\tilde{\delta}_{Q}  \right] , \\
&\eta_{2}=\frac{1 + 2\tilde{\epsilon}}{1 + 2\tilde{\delta}}
\left[
r_{0} + \frac{r_{1}}{1 + 2\tilde{\delta}} + \frac{r_{2}}{(1 + 2\tilde{\delta})^2}  
\right] ,
\end{align}
with
\begin{align}
&r_{0} = \tilde{\epsilon}_{Q}^2 , \\
&r_{1}=- \zeta_{Q} - 2  \tilde{\epsilon}_{Q} \, \tilde{\delta}_{Q}, \\
&r_{2}=2\tilde{\delta}_{Q}^2 . 
\end{align}
The parameter $\eta$ controls (along with the zero-dip normal-moveout velocity) all P-wave time-domain signatures for laterally homogeneous VTI media above a horizontal or dipping target reflector \cite[]{alkhalifah:1995,tsvankin:2001}. 
\subsection{Attenuation parameters}
The following Thomsen-type attenuation parameters are expressed directly through the elements $Q_{ij}$ and, therefore, are frequency-independent in constant-$Q$ VTI media: 
\begin{align} 
\label{eq:Ap0f}
&\mathcal{A}_{P0} = \tilde{\mathcal{A}}_{P0} ,  \\
\label{eq:As0f}
&\mathcal{A}_{S0} = \tilde{\mathcal{A}}_{S0} ,  \\
&\epsilon_{Q} = \tilde{\epsilon}_{Q} ,  \\
&\gamma_{Q} = \tilde{\gamma}_{Q} .
\end{align}

The attenuation parameter $\delta_{Q}$, however, also depends on the coefficients $M_{ij}^R$ (equation \ref{eq:deltaQdef}), which vary with frequency. The second-order approximation for $\delta_{Q}$ with respect to $\ln{| f/f_{0} |} $ is: 
\begin{equation} \label{eq:deltaQf}
\delta_{Q} =  \tilde{\delta}_{Q} + \frac{2}{\pi} \, Q_{33}^{-1} 
\, \zeta_{Q} \ln{\bigg| \frac{f}{f_{0}} \bigg|} \\
+ \frac{2}{\pi^2} \, Q_{33}^{-2} \, \xi_{Q} \ln^2{\bigg| \frac{f}{f_{0}} \bigg|} ,
\end{equation}
where $\zeta_{Q}$ is defined in equation \ref{eq:zetaQ}, and 
\begin{equation} \label{eq:xiQ}
\xi_{Q} = s_{0} (1-g_{Q})^3  + s_{1} (1-g_{Q})^2 \, \tilde{\delta}_{Q}  + s_{2}  (1-g_{Q}) \, \tilde{\delta}_{Q}^2 + s_{3} \, \tilde{\delta}_{Q}^3  .
\end{equation}
The explicit expressions for the coefficients $s_{i}$ are given in Appendix B. 

Because for $Q_{33} \gg 1$ the influence of the second-order term in equation \ref{eq:deltaQf} is insignificant,  the frequency variation of $\delta_{Q}$ is largely controlled by the coefficient $\zeta_{Q}$. For $\zeta_{Q}>0$, $\delta_{Q}$ monotonically increases with frequency. As follows from equations \ref{eq:zetaQ} and \ref{eq:d0}--\ref{eq:chi}, $\zeta_{Q}$ is a function of $g$ (equation \ref{eq:g}), $g_{Q}$ (equation \ref{eq:gQ}), $\tilde{\delta}$, and $\tilde{\delta}_Q$. 

\subsection{Numerical analysis}
Here, we analyze the above expressions for the Thomsen-type parameters numerically. The reference frequency is set as $f_{0}=40$~Hz and the frequency range as $[1, 200]$~Hz for all examples below. 

\begin{figure}
	\centering
	\subfloat[]{\includegraphics[width=2.2in]{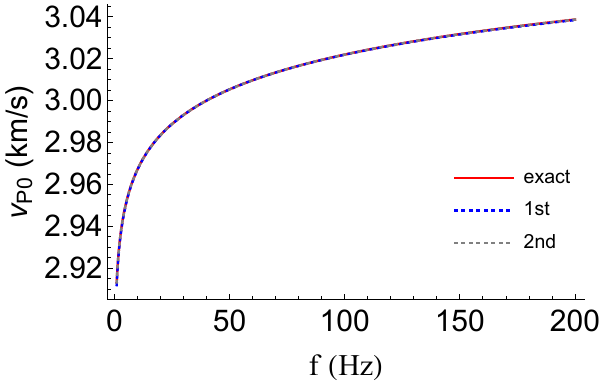}} \quad
	\subfloat[]{\includegraphics[width=2.2in]{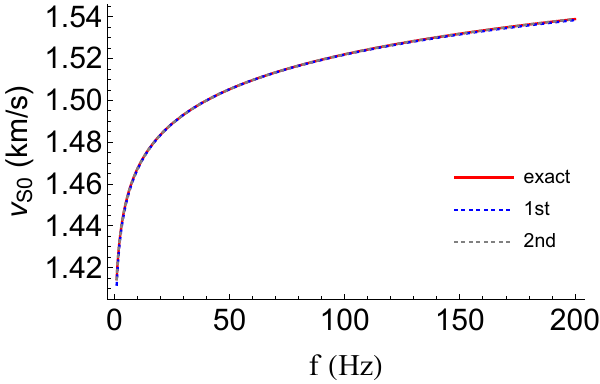}} 
	\caption{
		Frequency-dependent vertical velocities (a) $V_{P0}$ and (b) $V_{S0}$. ``Exact'' in the legend refers to the exact values, whereas ``1st'' and ``2nd'' denote the first- and second-order approximations with respect to $\ln{| f/f_{0} |}$, respectively. On plot (a),  $\tilde{V}_{P0}=3.0$~km/s and $Q_{33}=40$ ($\tilde{\mathcal{A}}_{P0}=0.0125$); on plot (b), $\tilde{V}_{S0}=1.5$~km/s and $Q_{55}=20$ ($\tilde{\mathcal{A}}_{S0}=0.025$). 
	}
	\label{fig:vp0vs0}
\end{figure}

First, we test the accuracy of the equations \ref{eq:vp0f} and \ref{eq:vs0f} for the vertical velocities and their first-order versions (i.e., those without the second-order term with respect to $\text{ln}|f/f_{0}|$). As demonstrated by Figure \ref{fig:vp0vs0}, the first-order approximations for $V_{P0}$ and $V_{S0}$ are sufficiently accurate even for strong attenuation in a wide frequency range. Overall, the frequency dependence of the vertical velocities is almost negligible, except for very low frequencies. 

\begin{table}
\centering
\caption{Medium parameters for two constant-$Q$ VTI models at the reference frequency $f_{0}=40$~Hz.}
\label{tab:tabl1}[H]
\begin{tabular}{c c c c c c c c c c c }
\toprule
\text{Model} & $\tilde{V}_{P0}$ & $\tilde{V}_{S0}$ &  $\tilde{\epsilon}$ &  $\tilde{\delta}$ & $\tilde{\gamma}$ & $\tilde{\mathcal{A}}_{P0}$ $(Q_{33})$ & $\tilde{\mathcal{A}}_{S0}$ $(Q_{55})$ & $\tilde{\epsilon}_{Q}$ & $\tilde{\delta}_{Q}$ & $\tilde{\gamma}_{Q}$ \\
\midrule
1 & 3.0  & 1.5  & 0.3 & -0.1  &  0.1    & 0.0125 (40)    & 0.0167 (30)    & -0.3      & -1.91     & 0.5    \\
2 & 3.0  & 1.5  & 0.3 & -0.1  &  0.2   & 0.0250 (20)    & 0.0333 (15)     & 0.3       & 0.98     & -0.2    \\
\bottomrule   
\end{tabular}
\end{table}

\begin{figure}[H]
	\centering
	\subfloat[]{\includegraphics[height=1.5in]{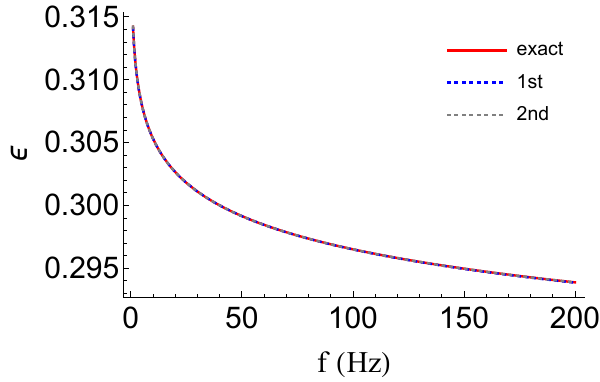}} \quad
	\subfloat[]{\includegraphics[height=1.5in]{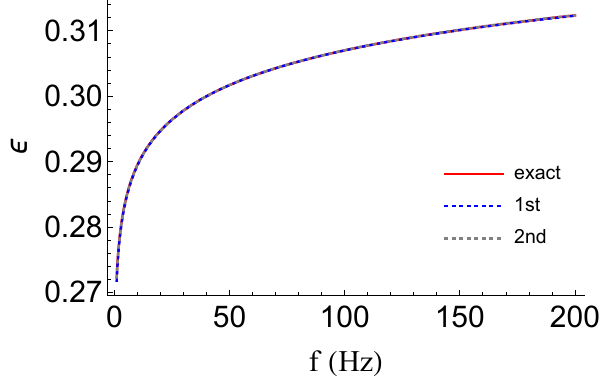}} 
	\caption{
		Variation of the Thomsen parameter $\epsilon$ with frequency for (a)~Model 1 and (b)~Model 2 from Table \ref{tab:tabl1}. The legend is the same as in Figure \ref{fig:vp0vs0}. 
	}
	\label{fig:thomsen1}
\end{figure}

\begin{figure}
	\centering
	\subfloat[]{\includegraphics[height=1.5in]{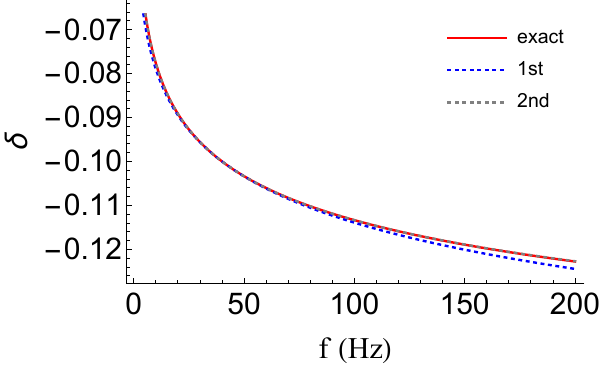}} \quad
	\subfloat[]{\includegraphics[height=1.5in]{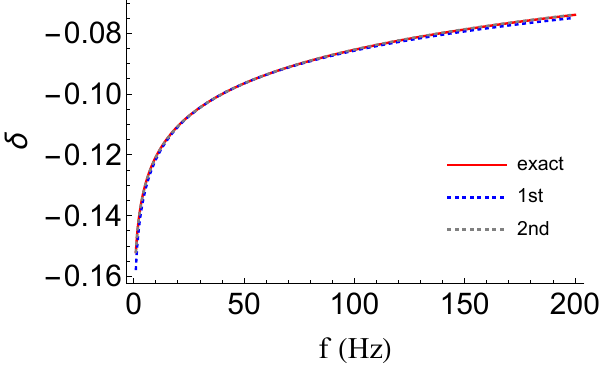}}  
	\caption{
		Same as Figure \ref{fig:thomsen1} but for the parameter $\delta$. 
	}
	\label{fig:thomsen2}
\end{figure}

\begin{figure}
	\centering
	\subfloat[]{\includegraphics[height=1.5in]{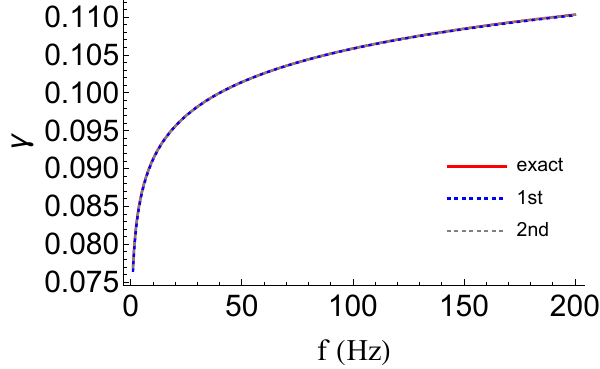}} \quad
	\subfloat[]{\includegraphics[height=1.5in]{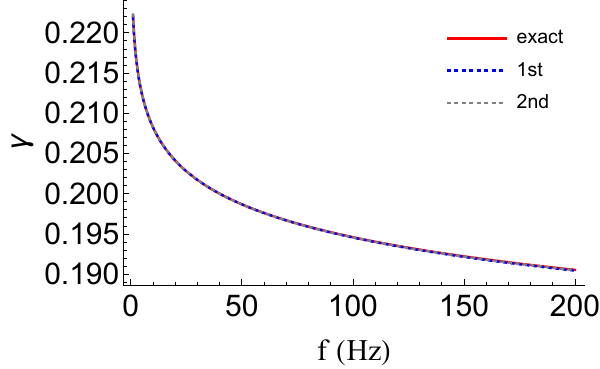}} 
	\caption{
		Same as Figure \ref{fig:thomsen1} but for the parameter $\gamma$. 
	}
	\label{fig:thomsen3}
\end{figure}

Figures \ref{fig:thomsen1}, \ref{fig:thomsen2} and \ref{fig:thomsen3} show that the first-order versions of equations \ref{eq:epsf}--\ref{eq:gamf} can accurately describe the variations of the anisotropy parameters $\epsilon$, $\delta$, and $\gamma$ with frequency. Comparison of Figures \ref{fig:thomsen1}, \ref{fig:thomsen2}, and \ref{fig:thomsen3} confirms that the reference parameters  $\tilde{\epsilon}_{Q}$, $\tilde{\delta}_{Q}$, and $\tilde{\gamma}_{Q}$ govern the frequency dependence of $\epsilon$, $\delta$, and $\gamma$. For example,  if $\tilde{\epsilon}_{Q}>0$, $\epsilon$ increases with frequency. As is the case for $V_{P0}$ and $V_{S0}$, the anisotropy coefficients vary with frequency primarily in the low-frequency range. \\~\\

\begin{figure}
	\centering
	\subfloat[]{\includegraphics[height=1.5in]{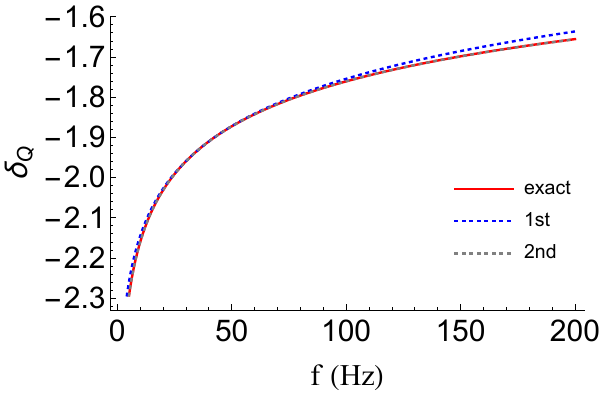}} \quad
	\subfloat[]{\includegraphics[height=1.5in]{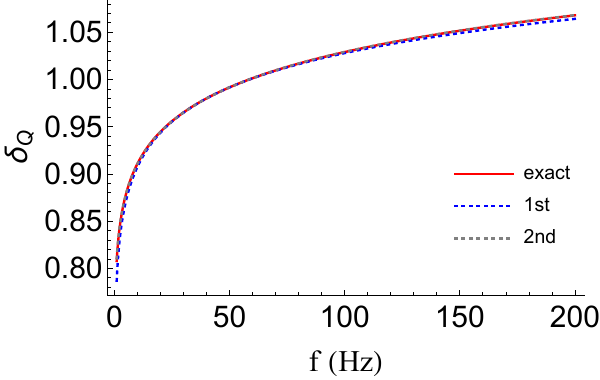}} 
	\caption{
		Frequency-dependent Thomsen-type attenuation parameter $\delta_{Q}$ for (a) Model 1 and (b) Model 2 from Table \ref{tab:tabl1}. The legend is the same as in Figure \ref{fig:vp0vs0}. 
	}
	\label{fig:deltaQ}
\end{figure}

Next, we investigate the only frequency-dependent attenuation-anisotropy parameter, $\delta_{Q}$, by comparing the exact equation for $\delta_{Q}$ with its first- and second-order approximations. 
The first-order equation accurately models $\delta_{Q}$ in a wide frequency range, whereas contribution of the second-order term is practically negligible (Figure \ref{fig:deltaQ}).  

\begin{figure}[H]
	\centering
	\subfloat[]{\includegraphics[height=2.2in]{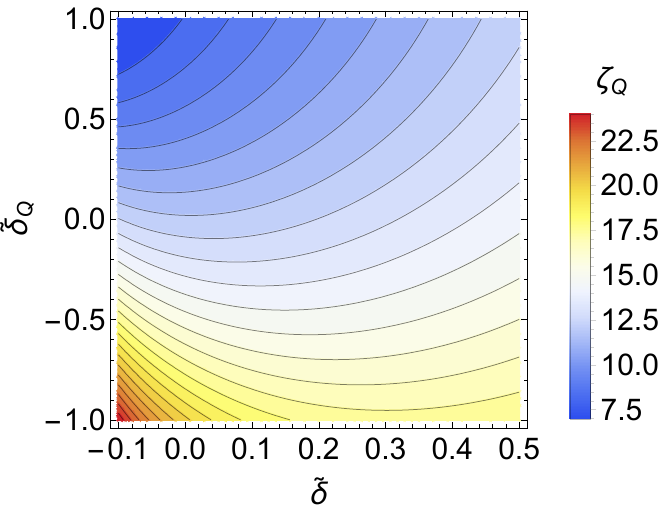}} \quad
	\subfloat[]{\includegraphics[height=2.2in]{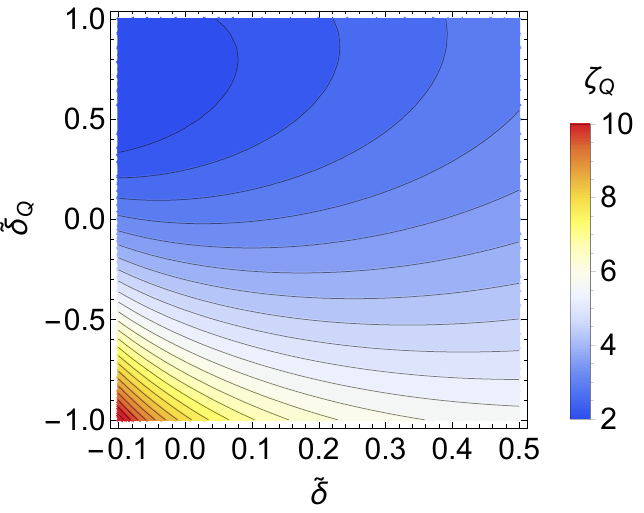}}  \\
	\subfloat[]{\includegraphics[height=2.2in]{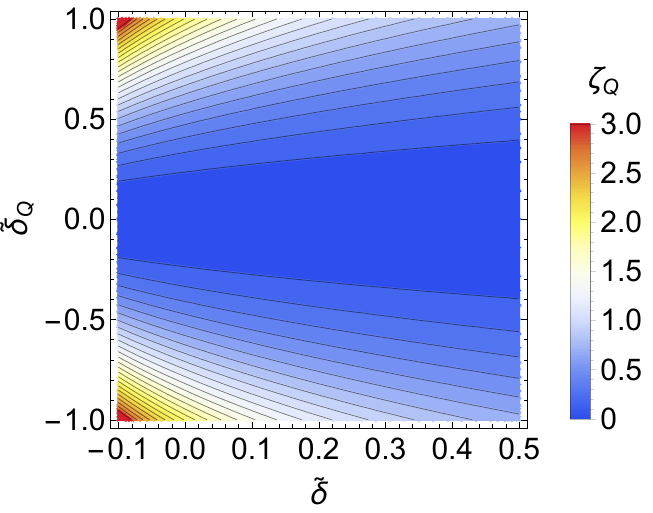}} \quad
	\subfloat[]{\includegraphics[height=2.2in]{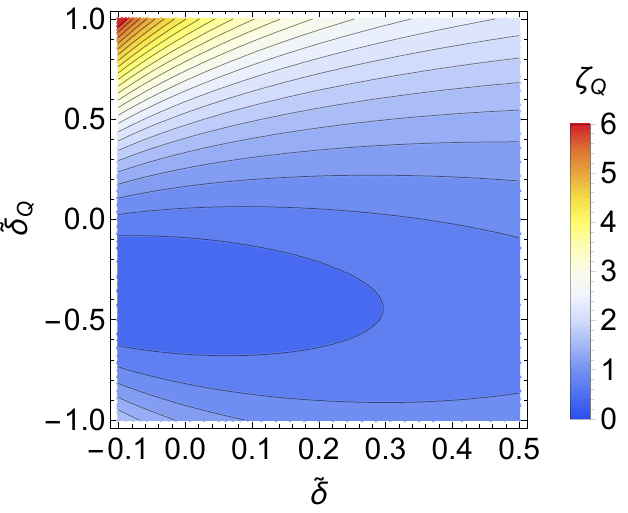}} 
	\caption{
		Contour plots of the coefficient $\zeta_{Q}$ as a function of $\tilde{\delta}$ and $\tilde{\delta}_{Q}$. The parameter $g=\tilde{V}_{S0}^2/\tilde{V}_{P0}^2=0.3$. The parameter $g_{Q}=Q_{55}^{-1}/Q_{33}^{-1}$ is set as (a) $g_{Q}=3$, (b) $g_{Q}=2$, (c) $g_{Q}=1$, and (d) $g_{Q}=0.5$. 
	}
	\label{fig:zetaQ}
\end{figure}

\begin{figure}
	\centering
	\subfloat[]{\includegraphics[width=2.2in]{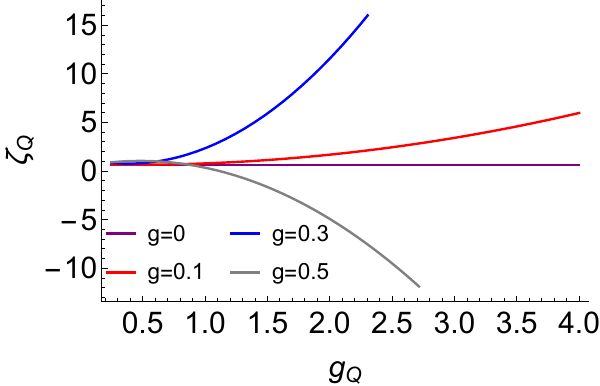}} \quad
	\subfloat[]{\includegraphics[width=2.2in]{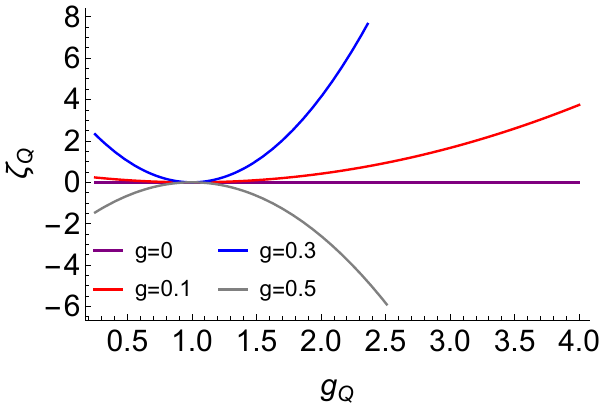}}  \\
	\subfloat[]{\includegraphics[width=2.2in]{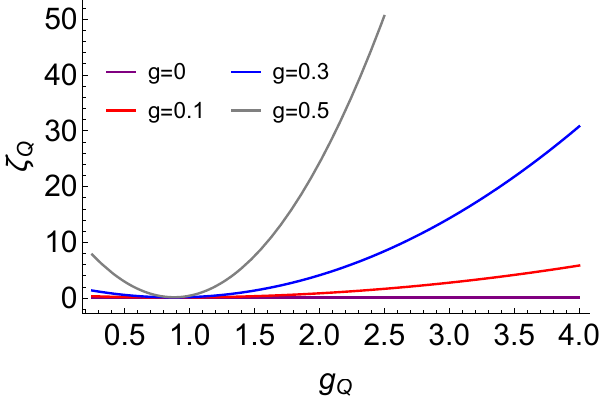}} \quad
	\subfloat[]{\includegraphics[width=2.2in]{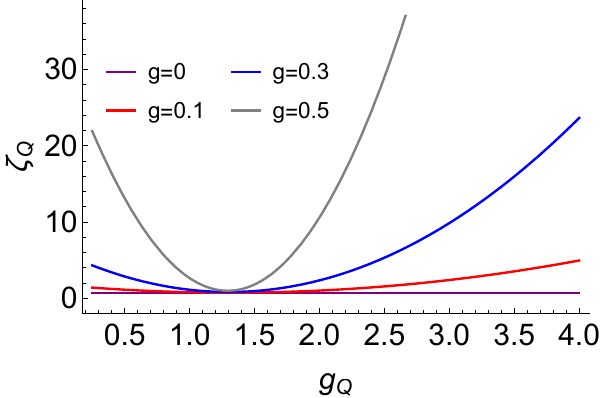}} 
	\caption{
		Variation of the coefficient $\zeta_{Q}$ with $g_{Q}$ for different values of $g$. (a) $\tilde{\delta}=-0.2$ and $\tilde{\delta}_{Q}=-0.6$;  (b) $\tilde{\delta}=-0.2$ and $\tilde{\delta}_{Q}=0$; (c) $\tilde{\delta}=0.2$ and $\tilde{\delta}_{Q}=-0.4$;  (d) $\tilde{\delta}=0.2$ and $\tilde{\delta}_{Q}=0.98$.  
	}
	\label{fig:zetaQgQ}
\end{figure}

\begin{figure}
	\centering
	\subfloat[]{\includegraphics[width=2.2in]{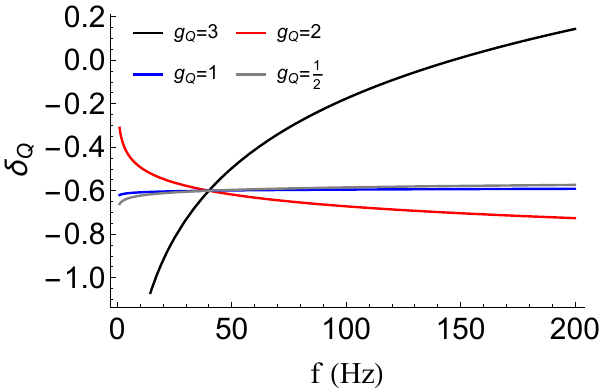}} \quad
	\subfloat[]{\includegraphics[width=2.2in]{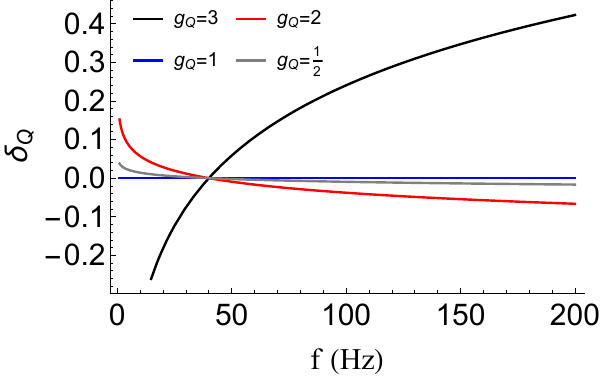}}  \\
	\subfloat[]{\includegraphics[width=2.2in]{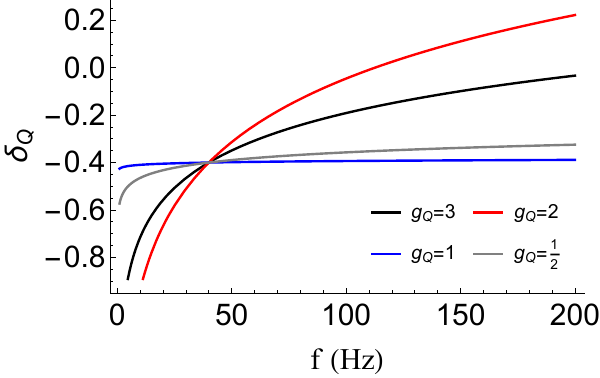}} \quad
	\subfloat[]{\includegraphics[width=2.2in]{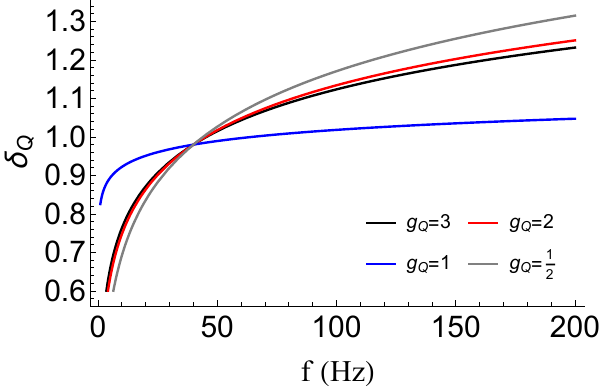}} 
	\caption{
		Variation of the attenuation parameter $\delta_{Q}$ with frequency for different $g_{Q}$ and $g=0.3$. The parameters $\tilde{\delta}$ and $\tilde{\delta}_{Q}$ are the same as in Figure \ref{fig:zetaQgQ}. 
	}
	\label{fig:delQgQ}
\end{figure}

As mentioned above, the coefficient $\zeta_{Q}$ in equation \ref{eq:deltaQf} is largely responsible for the frequency variation of $\delta_{Q}$ for a specified value of $Q_{33}$. Equation \ref{eq:zetaQ} shows that $\zeta_{Q}$ is a function of the parameters $g=\tilde{V}_{S0}^2/\tilde{V}_{P0}^2$, $g_{Q}=Q_{55}^{-1}/Q_{33}^{-1}$, $\tilde{\delta}$ and $\tilde{\delta}_{Q}$. Using the results from Figure \ref{fig:gQ_rocks}, we restrict $g_{Q}$ to the range $0.5 \le g_{Q} \le 3$. 
Figures \ref{fig:zetaQ} and \ref{fig:zetaQgQ} show that the smallest absolute value of $\zeta_{Q}$ corresponds to $g_{Q}=1$, and $|\zeta_{Q}|$ increases with the deviation of $g_{Q}$ from unity. As a result, the parameter $\delta_{Q}$ is almost independent of frequency for $g_{Q}=1$ (Figure \ref{fig:delQgQ}). 
Overall, the frequency dependence of $\delta_{Q}$ becomes noticeable for large $|g_{Q}-1|$ (e.g., $g_{Q}=3$; Figure \ref{fig:delQgQ}), but it is also influenced by the parameters $\tilde{\delta}$  and $\tilde{\delta}_{Q}$.  
For the most common values of $g_{Q}$ considered here, the parameter $\delta_{Q}$ significantly varies with $f$ only for low frequencies.  \\~\\

\section{Viscoacoustic constant-Q transverse isotropy}
\subsection{Simplified parameter expressions}
Next, we consider the so-called ``viscoacoustic'' constant-$Q$ media described by the Thomsen-type notation. The acoustic approximation is implemented by setting $\tilde{V}_{S0}=\mathcal{A}_{S0}=0$ in equations \ref{eq:deltaf}, \ref{eq:deltaQf}, and \ref{eq:etaf} \cite[]{HaoA2017a,HaoA2017b,hao.alkhalifah:2019} . The parameters $\delta$, $\eta$, and $\delta_Q$ then reduce to:
\begin{equation} \label{eq:deltafAcous}
\delta = \tilde{\delta} 
+ \frac{1}{\pi} \, Q_{33}^{-1} \, \tilde{\delta}_{Q} \ln{\bigg| \frac{f}{f_{0}} \bigg|}
+  \frac{1}{\pi^2} \, Q_{33}^{-2} \, \frac{ \tilde{\delta}_{Q}^2 }{1+2\tilde{\delta}}
\ln^2{\bigg| \frac{f}{f_{0}} \bigg|} ,
\end{equation}
\begin{equation} \label{eq:etaAcous}
\eta = \eta_{0} +  \eta_{1} \, Q_{33}^{-1} \, \ln{\bigg| \frac{f}{f_{0}} \bigg|} + 
 \left( \tilde{\epsilon}_{Q} - \frac{\tilde{\delta}_{Q}}{1 + 2\tilde{\delta}} \right) \eta_{1}
\, Q_{33}^{-2} \, \ln^2{\bigg| \frac{f}{f_{0}} \bigg|} ,
\end{equation}
\begin{equation} \label{eq:deltaQAcous}
\delta_{Q} = \tilde{\delta}_{Q} + \frac{2}{\pi} \, Q_{33}^{-1} \,
\frac{ \tilde{\delta}_{Q}^2 }{1+2\tilde{\delta}}  
\ln{\bigg| \frac{f}{f_{0}} \bigg|} + 
\frac{2}{\pi^2} \, Q_{33}^{-2} \,
\frac{ \tilde{\delta}_{Q}^3 }{(1+2\tilde{\delta})^2}
\ln^2{\bigg| \frac{f}{f_{0}} \bigg|} .
\end{equation}
Setting $\eta$ (equation \ref{eq:etaAcous}) to zero, which requires $\eta_{0}=\eta_{1}=0$ (see equations \ref{eq:eta0} and \ref{eq:eta1}), we obtain the elliptical conditions:
\begin{align}
\label{eq:cond_eq1}
&\tilde{\epsilon} = \tilde{\delta} , \\
\label{eq:cond_eq2}
&\tilde{\epsilon}_{Q} = \frac{ \tilde{\delta}_{Q} }{ 1 + 2 \tilde{\delta}} . 
\end{align}
Equations \ref{eq:cond_eq1} and \ref{eq:cond_eq2} make the parameters of viscoacoustic constant-$Q$ media satisfy the same conditions at all frequencies: 
\begin{align}
\label{eq:result_eq1}
&\epsilon = \delta , \\
\label{eq:result_eq2}
&\epsilon_{Q} = \frac{ \delta_{Q} }{ 1 + 2 \delta} ,
\end{align}
which follows from equations \ref{eq:epsf}, \ref{eq:eta1}, \ref{eq:deltafAcous}, and \ref{eq:deltaQAcous}. Equation \ref{eq:result_eq1} implies that the elliptical conditions at the reference frequency ensure that $\eta=0$ at all frequencies. 

For viscoelastic constant-$Q$ media discussed earlier, equation \ref{eq:result_eq1} remains approximately valid (i.e., the model is elliptical at all frequencies), if equations \ref{eq:cond_eq1} and \ref{eq:cond_eq2} are satisfied (see equations \ref{eq:etaf}--\ref{eq:eta1}). 

\subsection{Numerical validation}
Here, we verify the elliptical conditions (equations \ref{eq:cond_eq1} and \ref{eq:cond_eq2}) by computing the anellipticity parameter $\eta$. 
The exact $\eta$ is calculated using equations \ref{eq:epsdef}, \ref{eq:deltadef}, \ref{eq:epsQdef} and \ref{eq:deltaQdef} along with equations \ref{eq:Mkjar} and \ref{eq:gamij} under the acoustic approximation ($\tilde{V}_{S0}=0$ and $Q_{55}^{-1}=0$). The first-order approximation for $\eta$ is given by equation \ref{eq:etaAcous} without the second-order term with respect to $\text{ln}|f/f_{0}|$.  

\begin{figure}
	\centering
	\subfloat[]{\includegraphics[width=2.8in]{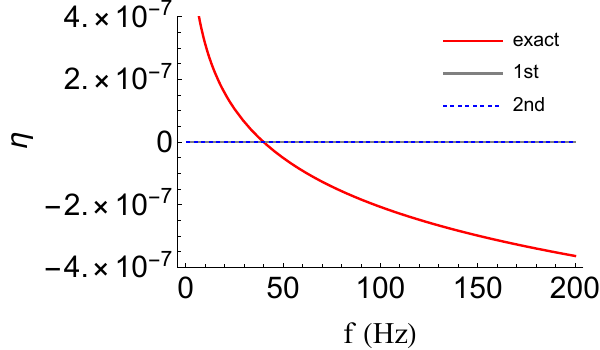}}
	\subfloat[]{\includegraphics[width=2.8in]{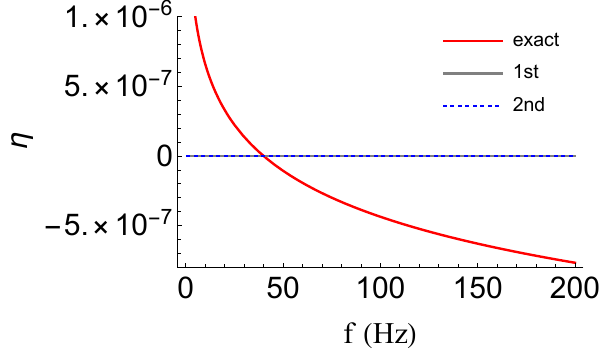}}  
	\caption{
		Variation of the anellipticity parameter $\eta$ with frequency under the elliptical conditions (equations \ref{eq:cond_eq1} and \ref{eq:cond_eq2}). The P-wave quality factor and reference vertical velocity at $f_{0}=40$~Hz are $Q_{33}=40$ and $\tilde{V}_{P0}=3$~km/s. The parameters $\tilde{\epsilon}$ and $\tilde{\epsilon}_{Q}$ are (a)~$\tilde{\epsilon}=0.3$ and $\tilde{\epsilon}_{Q}=-0.33$; (b)~$\tilde{\epsilon}=0.2$ and $\tilde{\epsilon}_{Q}=0.4$.
	}
	\label{fig:eta_ellip_acous}
\end{figure}

Figure \ref{fig:eta_ellip_acous} shows that for models that satisfy equations \ref{eq:cond_eq1}  and \ref{eq:cond_eq2} the exact anellipticity parameter is negligibly small for all frequencies (on the order of $10^{-7}$ for both models), which confirms that the elliptical conditions at the reference frequency lead to equation \ref{eq:result_eq1}. In addition, our testing confirms that the difference between the left and right sides of equation \ref{eq:result_eq2} is negligible, if equations \ref{eq:cond_eq1} and \ref{eq:cond_eq2} are satisfied.  

\section{Plane-wave attenuation in constant-Q  VTI media}
In this section, we apply the obtained expressions for the Thomsen-type parameters to study the normalized plane-wave attenuation coefficients in constant-$Q$ VTI media. The normalized phase attenuation coefficient is defined as $\mathcal{A} \equiv |\mathbf{k}_{I}|/|\mathbf{k}_{R}|$, where $\mathbf{k}_{R}$ and $\mathbf{k}_{I}$ denote the real and imaginary parts of the complex wave vector \cite[]{zhu.tsvankin:2006}. The words ``phase'' and ``normalized'' are omitted below for brevity.  The angle between $\mathbf{k}_{R}$ and $\mathbf{k}_{I}$ is called the ``inhomogeneity'' angle, which is not defined in plane-wave propagation (i.e., it is a free parameter that can vary within certain bounds). The coefficient $\mathcal{A}$ corresponding to $\mathbf{k}_{R} \parallel \mathbf{k}_{I}$ is approximately equal to the group attenuation coefficient, which can be estimated from seismic data, for a wide range of ``inhomogeneity'' angles \cite[]{behura:2009b,tsvankin.grechka:2011}. 

\subsection{Attenuation coefficients}
\cite{zhu.tsvankin:2006} and \cite{tsvankin.grechka:2011} show that the approximate attenuation coefficients of plane waves in viscoelastic constant-$Q$ VTI media are given by:
\begin{align}
\label{eq:Ap}
&\mathcal{A}_{P} = \mathcal{A}_{P0} \, (1 + \delta_{Q} \sin^2\theta \cos^2\theta  + \epsilon_{Q} \sin^4\theta ) ,  \\
\label{eq:Asv}
&\mathcal{A}_{SV} = \mathcal{A}_{S0} \, ( 1 + \sigma_{Q} \sin^2\theta \cos^2\theta ),  \\
\label{eq:Ash}
&\mathcal{A}_{SH} = \mathcal{A}_{S0} \, ( 1 + \gamma_{Q} \sin^2\theta  )  ,
\end{align}
where  the subscripts P, SV, and SH denote the wave types, and $\theta$ is the phase angle measured from the vertical. The quantity $\sigma_{Q}$ in equation \ref{eq:sigmaQ} is defined as \cite[]{zhu.tsvankin:2006}:
\begin{equation} \label{eq:sigmaQ}
\sigma_{Q} = 
2\frac{V_{P0}^2}{ V_{S0}^2} \left( \frac{Q_{55}}{Q_{33}} - 1 \right) (\epsilon-\delta) + 
\frac{V_{P0}^2 \, Q_{55} }{ V_{S0}^2 \, Q_{33}  } (\epsilon_{Q}-\delta_{Q} ) . 
\end{equation}

Equations \ref{eq:Ap}--\ref{eq:Ash} are derived under the assumption of weak attenuation and weak anisotropy (in both velocity and attenuation). Note that the effective quality factor, assumed to be frequency-independent in constant-$Q$ TI media, is proportional to the inverse of the attenuation coefficient \cite[]{zhu.tsvankin:2006}. 

Substitution of the Thomsen parameters from equations \ref{eq:vp0f}--\ref{eq:gamf} and \ref{eq:Ap0f}--\ref{eq:deltaQf} into equations \ref{eq:Ap}--\ref{eq:sigmaQ} allows us to separate the frequency-dependent parts of the attenuation coefficients. 
The approximate P-wave attenuation coefficient then becomes (only the linear term in  $\ln{| f/f_{0} |}$ is retained):  
\begin{equation} \label{eq:Qpf}
\mathcal{A}_{P} =\tilde{\mathcal{A}} _{P0} \left( 1 + \tilde{\delta}_{Q} \sin^2\theta \cos^2\theta  + \tilde{\epsilon}_{Q} \, \sin^4\theta 
+ R_{P} \ln{\bigg| \frac{f}{f_{0}} \bigg|} \right)   ,
\end{equation}
where $R_{P}$ controls the derivative of  $\mathcal{A}_{P} $ with respect to $\ln |f/f_{0}|$, 
\begin{equation} \label{eq:Rp}
R_{P} =\frac{1}{\pi} \, \tilde{\mathcal{A}}_{P0} \, \zeta_{Q} \sin^2(2\theta) ;
\end{equation}
$\zeta_{Q}$ is defined in equation \ref{eq:zetaQ}. 

For SV-waves, 
\begin{equation} \label{eq:Qsvf} 
\mathcal{A}_{SV} = \tilde{\mathcal{A}}_{S0} \left( 1 + \tilde{\sigma}_{Q} \sin^2\theta \cos^2\theta 
+ R_{SV}  \ln{\bigg| \frac{f}{f_{0}} \bigg|} \right) ,
\end{equation}
with
\begin{equation}
\tilde{\sigma}_{Q} = 
2\tilde{\sigma} \left( g_{Q} - 1 \right) + 
\frac{1 }{ g \, g_{Q} } (\tilde{\epsilon}_{Q}-\tilde{\delta}_{Q} ) , 
\end{equation}
\begin{equation} \label{eq:Rsv}
R_{SV} = \frac{1}{\pi} \, \tilde{\mathcal{A}}_{S0} \, \sigma'_{Q}  \sin^2(2\theta)   ,
\end{equation}
\begin{equation}
\sigma'_{Q} =  \frac{2(1-g_{Q})}{g \, g_{Q}^2} \left[
(1-g_{Q}) (\tilde{\epsilon}-\tilde{\delta})
-\tilde{\delta}_Q + (1 + \tilde{\epsilon})\tilde{\epsilon}_Q
\right]
- \frac{ \zeta_{Q}}{g \, g_{Q}^2} , 
\end{equation}
where $g$ and $g_{Q}$ are given by equations \ref{eq:g} and \ref{eq:gQ}, respectively. The factor $R_{SV}$ controls the derivative of $\mathcal{A}_{SV}$ with respect to $\ln |f/f_{0}|$. 

The terms $\tilde{\mathcal{A}} _{P0} \, R_{P}$ and $\tilde{\mathcal{A}}_{S0} \, R_{SV}$ define the rate of the P- and SV-wave attenuation-coefficient change (increase or decrease) with respect to $\ln |f/f_{0}|$. The larger  $R_{P}$ and $R_{SV}$ are, the stronger is the dispersion (frequency dependence) of $\mathcal{A}_{P}$ and $\mathcal{A}_{SV}$. Therefore, $R_{P}$ and $R_{SV}$ can be called the P- and SV-wave dispersion factors, respectively.   

The SH-wave attenuation coefficient (equation \ref{eq:Ash}) is independent of frequency, with $\gamma_{Q}=\tilde{\gamma}_{Q}$:
\begin{equation} \label{eq:Qsh}
\mathcal{A}_{SH} = \tilde{\mathcal{A}}_{S0} (1 + \tilde{\gamma}_{Q} \sin^2\theta ) .
\end{equation}

\subsection{Numerical dispersion analysis}
\begin{figure}
\vspace{-13ex}
	\centering
	\subfloat[]{\includegraphics[width=2.2in]{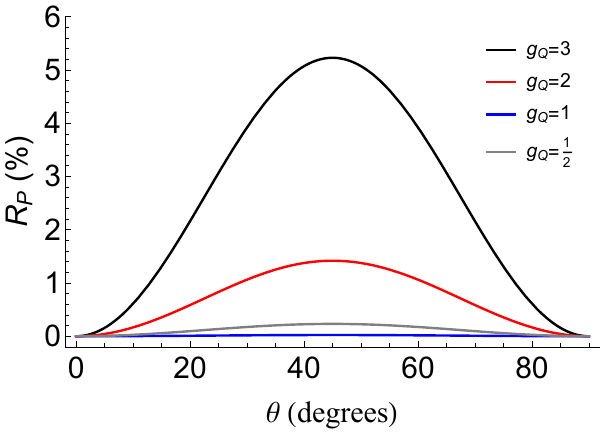}}  
	\subfloat[]{\includegraphics[width=2.2in]{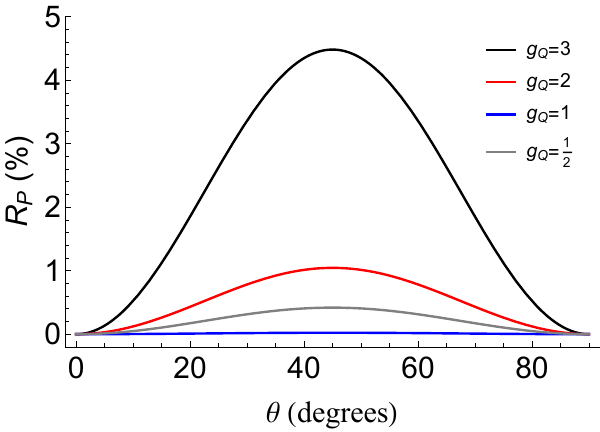}} \\  
	\subfloat[]{\includegraphics[width=2.2in]{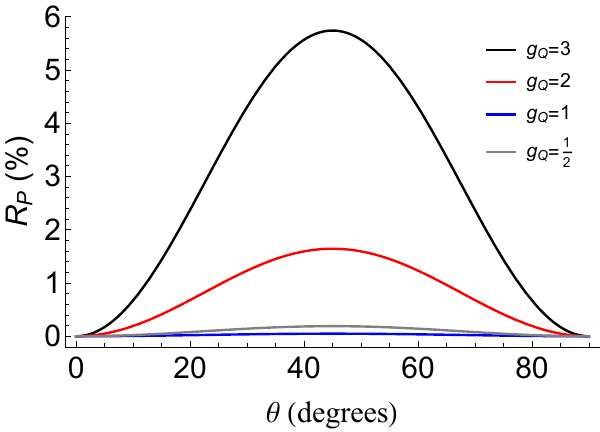}}  
	\subfloat[]{\label{fig:Rpd} \includegraphics[width=2.2in]{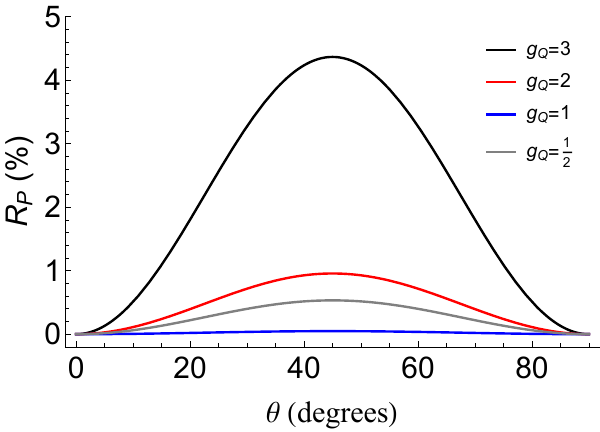}}  
	\caption{
    Variation of the P-wave dispersion factor $R_{P}$ (equation \ref{eq:Rp}) with the phase angle for different $g_{Q}$. The reference parameters defined at $f_{0}=40$~Hz are $\tilde{V}_{P0}=3.0$~km/s, $g=0.3$, $\tilde{\epsilon}=0.2$, $\tilde{\mathcal{A}}_{P0} = 0.0125$ (corresponding to $Q_{33}=40$), and $\tilde{\epsilon}_{Q}=-0.1$. (a) $\tilde{\delta}=0.1$ and $\tilde{\delta}_{Q}=-0.2$; (b) $\tilde{\delta}=0.1$ and $\tilde{\delta}_{Q}=0.2$; (c) $\tilde{\delta}=-0.1$ and $\tilde{\delta}_{Q}=-0.2$; (d) $\tilde{\delta}=-0.1$ and $\tilde{\delta}_{Q}=0.2$.   
	}
	\label{fig:Rp}
\end{figure}

\begin{figure}
	\centering
	\subfloat[]{\includegraphics[width=2.2in]{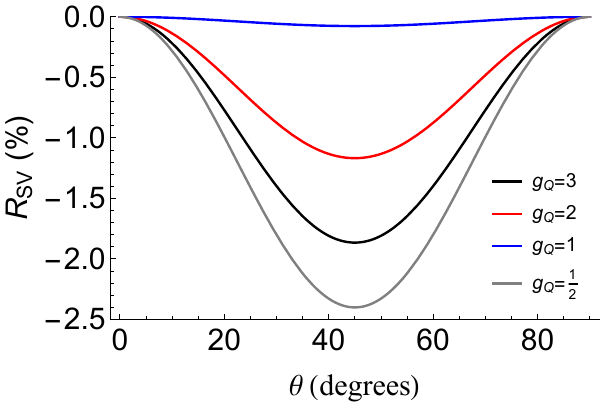}}  
	\subfloat[]{\includegraphics[width=2.2in]{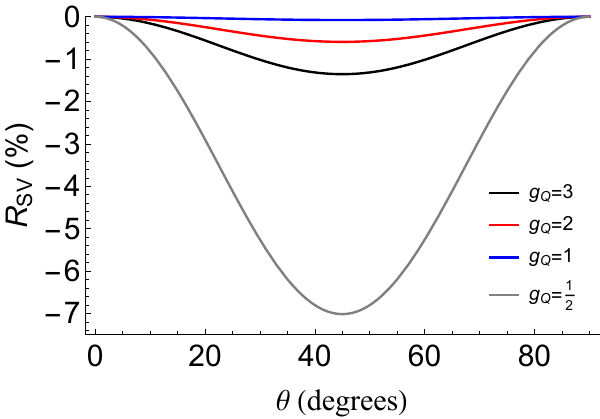}} \\  
	\subfloat[]{\includegraphics[width=2.2in]{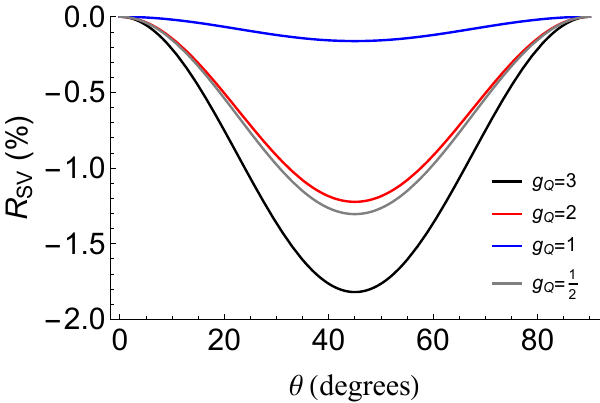}}  
	\subfloat[]{\label{fig:Rsvd} \includegraphics[width=2.2in]{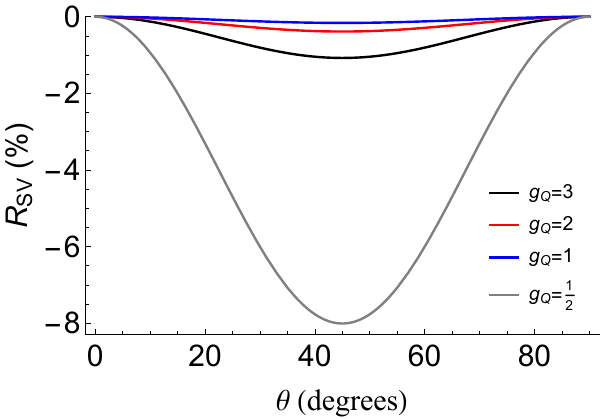}}  
	\caption{
    Variation of the SV-wave dispersion factor $R_{SV}$ (equation \ref{eq:Rsv}) with the phase angle for different $g_{Q}$. The reference parameters defined at $f_{0}=40$~Hz are $\tilde{V}_{P0}=3.0$~km/s, $g=0.3$, $\tilde{\epsilon}=0.2$, $\tilde{\mathcal{A}}_{S0} = 0.0125$ (corresponding to $Q_{55}=40$), $\tilde{\epsilon}_{Q}=-0.1$ and $\tilde{\epsilon}_{Q}=-0.2$. The parameters  $\tilde{\delta}$ and $\tilde{\delta}_{Q}$ are the same as in Figure \ref{fig:Rp}. 
	}
	\label{fig:Rsv}
\end{figure}

Here, we evaluate the frequency dependence of the attenuation coefficients of P- and SV-waves, starting with the dispersion factors $R_{P}$ and $R_{SV}$ (equations \ref{eq:Rp} and \ref{eq:Rsv}). As before, we restrict $g_{Q}$ to the realistic range $0.5 < g_{Q} \le 3$ (Figure \ref{fig:gQ_rocks}). 
Figures \ref{fig:Rp} and \ref{fig:Rsv} show that $g_{Q}=1$ yields the smallest values of $R_{P}$ and $R_{SV}$; the dispersion factors and the magnitude of their variation with angle increase with the deviation of $g_{Q}$ from unity. 

Next, we use the medium parameters from Figures \ref{fig:Rpd} and \ref{fig:Rsvd} to calculate the exact attenuation coefficients for P- and SV-waves (respectively) at three frequencies. For the reference frequency $f_{0}=40$~Hz, the term $\ln| f/f_{0}|$ in equations \ref{eq:Rp} and \ref{eq:Rsv} is close to $-1$ at $f=15$~Hz and $1$ at $f=109$~Hz. In agreement with equations \ref{eq:Qpf} and \ref{eq:Qsvf},  the variation of $\mathcal{A}_{P}$ with $\ln| f/f_{0}|$ between $15$~Hz and $40$~Hz (and $40$~Hz and $109$~Hz) is approximately proportional to $R_{P}$, and the corresponding variation of $\mathcal{A}_{SV}$ is approximately proportional to $R_{SV}$. 

\begin{figure}
	\centering
	\subfloat[]{\includegraphics[width=2.2in]{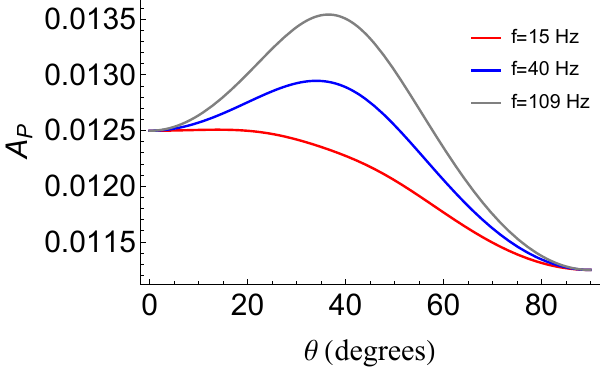}} \quad
	\subfloat[]{\includegraphics[width=2.2in]{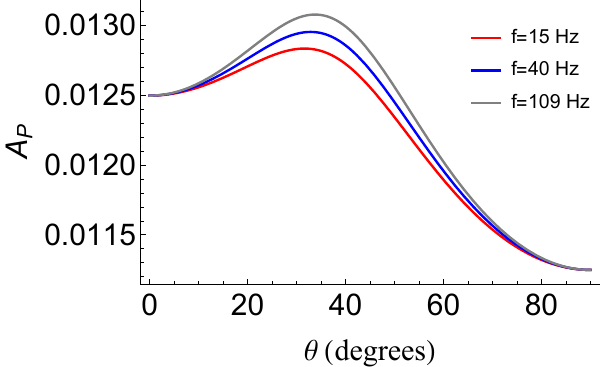}} \\
	\subfloat[]{\includegraphics[width=2.2in]{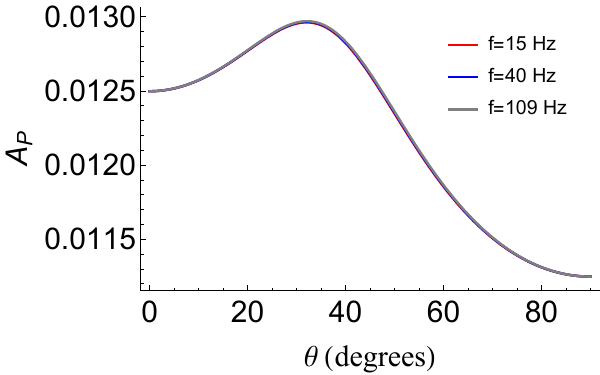}} \quad
	\subfloat[]{\includegraphics[width=2.2in]{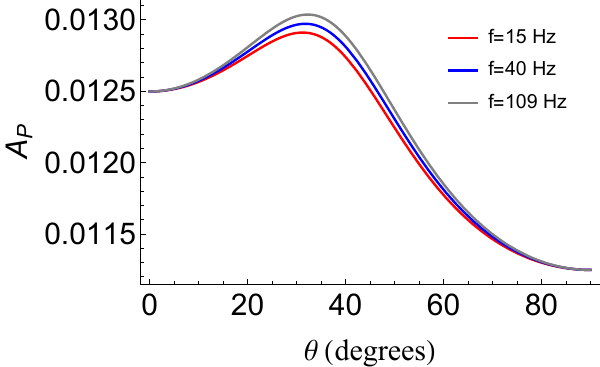}} 
	\caption{
		Variation of the P-wave normalized phase attenuation coefficient with the phase angle at different frequencies. The medium parameters are the same as in Figure \ref{fig:Rpd}, and (a) $g_{Q}=3$; (b) $g_{Q}=2$; (c) $g_{Q}=1$; (d) $g_{Q}=0.5$. 
	}
	\label{fig:Ap}
\end{figure}

\begin{figure}
	\centering
	\subfloat[]{\includegraphics[width=2.2in]{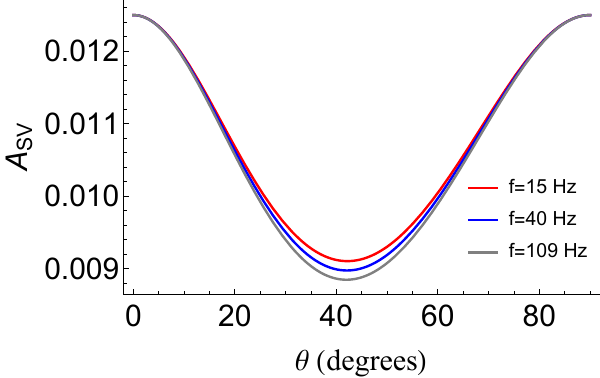}} \quad
	\subfloat[]{\includegraphics[width=2.2in]{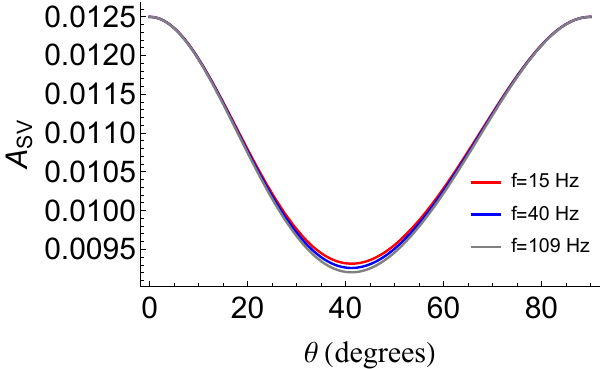}} \\
	\subfloat[]{\includegraphics[width=2.2in]{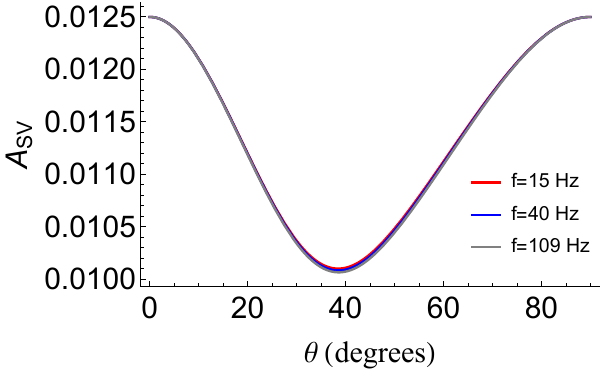}} \quad
	\subfloat[]{\includegraphics[width=2.2in]{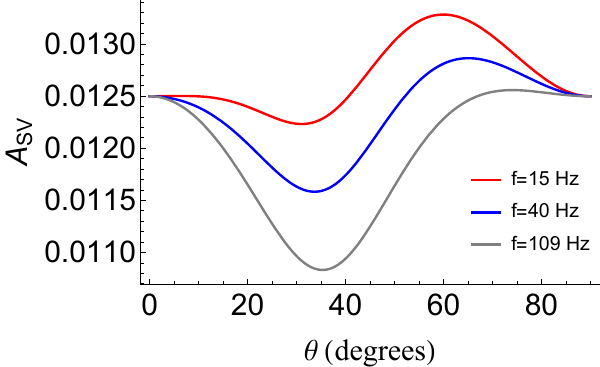}} 
	\caption{
		Variation of the SV-wave attenuation coefficient with the phase angle at different frequencies. The medium parameters are the same as in Figure \ref{fig:Rsvd}, and (a) $g_{Q}=3$; (b) $g_{Q}=2$; (c) $g_{Q}=1$; (d) $g_{Q}=0.5$.  
	}
	\label{fig:Asv}
\end{figure}

\begin{figure}
	\centering
    \subfloat[]{\includegraphics[width=2.8in]{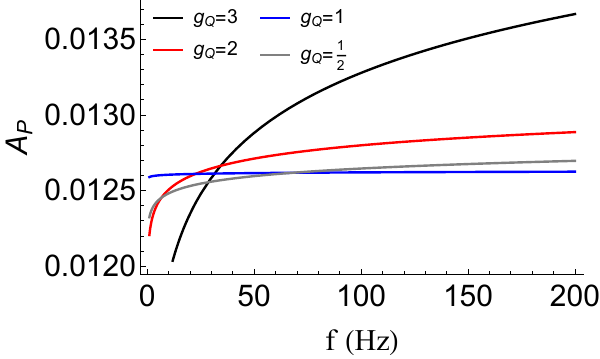}}
    \subfloat[]{\includegraphics[width=2.8in]{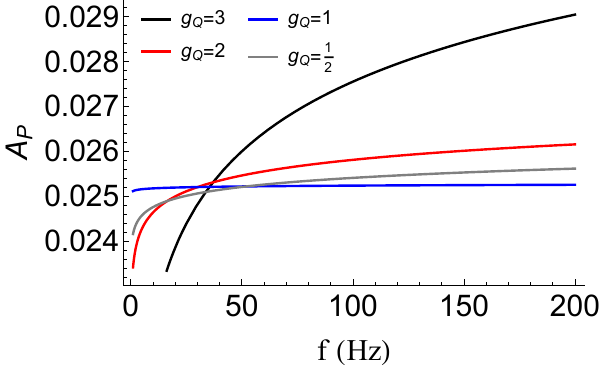}}
	\caption{
		Variation of the P-wave attenuation coefficient with frequency at $\theta=45\degree$ for different $g_{Q}$. Except for $\tilde{\mathcal{A}}_{P0}$, the medium parameters are the same as in Figures \ref{fig:Rpd} and \ref{fig:Ap}. On plot (a), $\tilde{\mathcal{A}}_{P0} = 0.0125$ (corresponding to $Q_{33}=40$); on plot (b), $\tilde{\mathcal{A}}_{P0} = 0.025$ (corresponding to $Q_{33}=20$). 
	}
	\label{fig:Ap45}
\end{figure}

\begin{figure}
	\centering
    \subfloat[]{\includegraphics[width=2.8in]{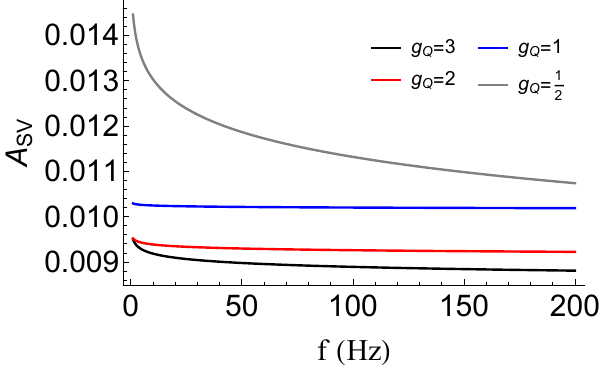}}
    \subfloat[]{\includegraphics[width=2.8in]{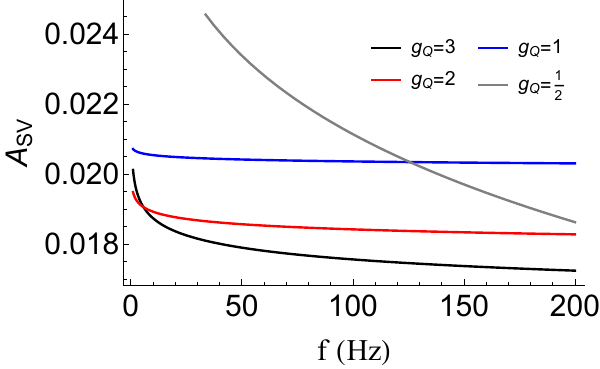}}
	\caption{
		Variation of the SV-wave attenuation coefficient with frequency at $\theta=45\degree$ for different $g_{Q}$. Except for $\tilde{\mathcal{A}}_{S0}$, the medium parameters are the same as in Figures \ref{fig:Rsvd} and \ref{fig:Asv}. On plot (a), $\tilde{\mathcal{A}}_{S0} = 0.0125$ (corresponding to $Q_{55}=40$); on plot (b), $\tilde{\mathcal{A}}_{S0} = 0.025$ (corresponding to $Q_{55}=20$). 
	}
	\label{fig:Asv45}
\end{figure}

Figures \ref{fig:Ap} and \ref{fig:Asv} show that the frequency dependence of the P- and SV-wave attenuation coefficients $\mathcal{A}_{P}$ and $\mathcal{A}_{SV}$ is generally mild. However, they may become noticeable for propagation angles close to $45\degree$ as illustrated in Figures \ref{fig:Ap45} and \ref{fig:Asv45}. Both $\mathcal{A}_{P}$ and $\mathcal{A}_{SV}$ exhibit a more significant variation with frequency for strongly attenuative media ($Q_{33}$=$Q_{55}$=20) when $g_{Q} \ge 2$ (for P-waves) and $g_{Q} \le 0.5$ (for SV-waves). 

\section{Conclusions}
We obtained concise analytic expressions for the Thomsen-type parameters of constant-$Q$ TI media. 
All Thomsen velocity parameters ($V_{P0}$, $V_{S0}$, $\epsilon$, $\delta$ and $\gamma$) are frequency dependent, with the reference attenuation parameters $\tilde{\mathcal{A}}_{P0}$ (proportional to $1/Q_{33}$) and $\tilde{\mathcal{A}}_{S0}$ (proportional to $1/Q_{55}$) controlling the dispersion (frequency dependence) of the vertical velocities $V_{P0}$ and $V_{S0}$, respectively. The reference attenuation parameters $\tilde{\epsilon}_{Q}$, $\tilde{\delta}_{Q}$, and $\tilde{\gamma}_{Q}$ govern the variations of the anisotropy parameters $\epsilon$, $\delta$, and $\gamma$ with frequency. However, the frequency dependence of all Thomsen velocity parameters is weak  in a wide frequency range, even for strong attenuation. In viscoacoustic constant-$Q$ TI media, the elliptical conditions at the reference frequency ensure that the anellipticity parameter $\eta$ vanishes for all frequencies. 

Despite the fact that all $Q_{ij}$ elements in constant-$Q$ TI media are frequency independent, one of the Thomsen-type attenuation parameters ($\delta_{Q}$) does vary with frequency. The frequency dependence of $\delta_{Q}$ is controlled by the newly defined coefficient $\zeta_{Q}$ and can be substantial when $\zeta_{Q}$ has a large magnitude. As a result, the frequency variation of the P- and SV-wave attenuation coefficients may be non-negligible at oblique propagation angles with the symmetry axis.
That variation is highly sensitive to the ratio of the vertical quality factors $g_{Q}=Q_{33}/Q_{55}$. Both attenuation coefficients are insensitive to frequency for $g_{Q}=1$, whereas their frequency dependence is most substantial for $g_{Q} \ge 3$ (for P-waves) and $g_{Q} \le 0.5$ (for SV-waves). 
In contrast, the SH-wave attenuation coefficient in constant-$Q$ TI media is frequency-independent. 

The constant-$Q$ assumption is often made in attenuation analysis because the effective attenuation coefficients estimated from seismic data (e.g., using the spectral-ratio method) become linear functions of frequency. However, our results show that this linear dependence may not hold for constant-$Q$ TI models, which can cause errors in the inversion for the attenuation parameters. 

\begin{appendix}
\section{Appendix A: Complex stiffness coefficients expressed in terms of the Thomsen-type parameters}
The stiffness coefficients for the constant-$Q$ dissipative VTI model (equations \ref{eq:Mgen}--\ref{eq:gamij}) can be found at the reference frequency as $M_{ij}|_{f=f_{0}}=\tilde{M}_{ij}^R(1-i/Q_{ij})$.  Using the parameter definitions in equations \ref{eq:vp0def}--\ref{eq:deltaQdef}, we express $\tilde{M}_{ij}^R$ and $Q_{ij}$ in terms of the reference Thomsen-type parameters as follows: 
\begin{equation}
\tilde{M}_{33}^R = \rho \tilde{V}_{P0}^2 , 
\end{equation}
\begin{equation}
\tilde{M}_{55}^R = \rho \tilde{V}_{S0}^2 , 
\end{equation}
\begin{equation}
\tilde{M}_{11}^R = \rho \tilde{V}_{P0}^2 (1 + 2\tilde{\epsilon}) , 
\end{equation}
\begin{equation}
\tilde{M}_{66}^R = \rho \tilde{V}_{S0}^2 (1 + 2\tilde{\gamma}) , 
\end{equation}
\begin{equation}
\tilde{M}_{13}^R = - \rho \tilde{V}_{S0}^2 + \rho \sqrt{(\tilde{V}_{P0}^2-\tilde{V}_{S0}^2) \left[ (1+2\tilde{\delta})\tilde{V}_{P0}^2-\tilde{V}_{S0}^2 \right] },
\end{equation}
\begin{equation}
Q_{33}^{-1}  =  \frac{2\tilde{\mathcal{A}}_{P0}}{1 - \tilde{\mathcal{A}}_{P0}^2}  ,
\end{equation}
\begin{equation}
Q_{55}^{-1}  =   \frac{2\tilde{\mathcal{A}}_{S0}}{1 - \tilde{\mathcal{A}}_{S0}^2}  ,
\end{equation}
\begin{equation}
Q_{11}^{-1}  = Q_{33}^{-1} (1 + \tilde{\epsilon}_{Q}) ,
\end{equation}
\begin{equation}
Q_{66}^{-1}  =  Q_{55}^{-1} (1 + \tilde{\gamma}_{Q}) 
\end{equation}
\begin{equation}
Q_{13}^{-1}  = \tilde{Q}_{33}^{-1} \left(1 +\tilde{\delta}_{Q} f_{1} + f_{2} \right) - Q_{55}^{-1} f_{2} ,
\end{equation}
with
\begin{equation}
f_{1} = \frac{\tilde{M}_{33}^R \,  (\tilde{M}_{33}^R-\tilde{M}_{55}^R)}{2\tilde{M}_{13}^R(\tilde{M}_{13}^R + \tilde{M}_{55}^R) } ,
\end{equation}
\begin{equation}
f_{2} = \frac{\tilde{M}_{55}^R \, (\tilde{M}_{13}^R + \tilde{M}_{33}^R)^2 }{2\tilde{M}_{13}^R (\tilde{M}_{13}^R + \tilde{M}_{55}^R)(\tilde{M}_{33}^R - \tilde{M}_{55}^R)} .
\end{equation}

%
\section{Appendix B: Explicit expressions for $\mathbf{s_{n}}$}
Here, we provide explicit expressions for the coefficients $s_{n}$ in equation \ref{eq:xiQ}. 

The coefficient $s_{0}$ is given by:
\begin{equation}
{s}_{0} = 
\frac{g (1-g+\chi)^2
(h_{0} + h_{1}g + h_{2}g^2 + h_{3}g^3 + h_{4}g^4 + h_{5}g^5 )
}{(1-g)^3 \chi ^3 (g-\chi )^2} ,
\end{equation}
where
\begin{align}
&h_{0} = -(1 + 2 \tilde{\delta})^2 \chi , \\
&h_{1}=(1 + 2 \tilde{\delta}) (5 + 10 \tilde{\delta} +2 \chi) , \\
&h_{2}=(1 + 2 \tilde{\delta})\left[ 2 (\tilde{\delta} -3) \chi -13 \tilde{\delta} -14 \right] , \\
&h_{3}=\tilde{\delta}  (7 \tilde{\delta} +9 \chi +30)+7 \chi +15 , \\
&h_{4}=-\tilde{\delta} ^2 - 2 (\tilde{\delta} +1) \chi  - 11 \tilde{\delta} - 8 ,  \\
&h_{5}=2 (1 + 2 \tilde{\delta}) .
\end{align}
For the coefficient $s_{1}$ we have:
\begin{equation}
s_{1} = \frac{3 g (g-\chi -1) (k_{0} + k_{1}g + k_{2}g^2 + k_{3}g^3 + k_{4}g^4)
}{2(1-g) \chi ^3 (g-\chi )^2} ,
\end{equation}
where
\begin{align}
&k_{0} = -2 (1 + 2 \tilde{\delta})^2 , \\
&k_{1}=2\left[1 +\chi + 4 \tilde{\delta}  (\tilde{\delta} +\chi +1) \right] , \\
&k_{2}= 2 (\chi +1)-\tilde{\delta}  (\chi +3) , \\
&k_{3}=-(\tilde{\delta} +2 \chi +4) , \\
&k_{4}= 2 ;
\end{align}
Finally, the coefficient $s_{2}$ has the form:
\begin{equation}
s_{2} = \frac{3 g\left[3 + 6 \tilde{\delta} + 2\chi  - 3 g (3\tilde{\delta} + 2\chi +3) +  3g^2 (\tilde{\delta} +\chi +3) -  3g^3 \right]}{2\chi ^3 (g-\chi )^2}  ;
\end{equation}
\begin{equation}
s_{3} = \frac{(1-g)^2 (4 \chi -3g)}{4 \chi ^3 (g-\chi )^2} .
\end{equation}
The quantities $g$ and $\chi$ are defined in equations \ref{eq:g} and \ref{eq:chi}, respectively. 

\end{appendix}


\bibliographystyle{./macros/elsarticle-num}
\bibliography{./refs/refs20220727,./refs/qi_refs20220719}

\end{document}